\documentclass[12pt,onecolumn]{IEEEtran}
\usepackage[colorlinks,bookmarksopen,bookmarksnumbered,citecolor=red,urlcolor=red,]{hyperref}
 \usepackage{CJK}
 \usepackage{indentfirst}
\usepackage{algorithm,algorithmic,amsbsy,amsmath,amssymb,epsfig,bbm,mathrsfs, bbm} 
\usepackage{multirow}
\usepackage{mathrsfs}
 \usepackage{epstopdf}

\usepackage{verbatim}
\usepackage{amsfonts}
\usepackage{amsthm}
\usepackage[subfigure]{graphfig}

\begin{document}

\title{SSthreshless Start: A Sender-Side TCP Intelligence for Long Fat Network}

\author {\authorblockN{~Xiao~Lu$^{\dagger}$, Ke~Zhang$^{\dagger}$, Chuan~Heng~Foh$^{\ddagger}$, and Cheng~Peng~Fu$^{\dagger}$\\
\authorblockA{~$^{\dagger}$ School of Computer Engineering,
        Nanyang Technological University, Singapore\\ $^{\ddagger}$ Department of Electrical Engineering, University of Surrey, UK } }}\maketitle

\maketitle

\begin{abstract}
Measurement shows that $85\%$ of TCP flows in the internet are short-lived
flows that stay most of their operation in the TCP startup phase. However, many
previous studies indicate that the traditional TCP Slow Start algorithm does
not perform well, especially in long fat networks. Two obvious problems are
known to impact the Slow Start performance, which are the blind initial setting
of the Slow Start threshold and the aggressive increase of the probing rate
during the startup phase regardless of the buffer sizes along the path. Current
efforts focusing on tuning the Slow Start threshold and/or probing rate during
the startup phase have not been considered very effective, which has prompted
an investigation with a different approach.

In this paper, we present a novel TCP startup method, called
\emph{threshold-less slow start} or SSthreshless Start, which does not need the
Slow Start threshold to operate. Instead, SSthreshless Start uses the backlog
status at bottleneck buffer to adaptively adjust probing rate which allows
better seizing of the available bandwidth. Comparing to the traditional and
other major modified startup methods, our simulation results show that
SSthreshless Start achieves significant performance improvement during the
startup phase. Moreover, SSthreshless Start scales well with a wide range of
buffer size, propagation delay and network bandwidth. Besides, it shows
excellent friendliness when operating simultaneously with the currently popular
TCP NewReno connections.
\end{abstract}

\section{Introduction}

TCP is a connection-oriented, reliable and in-order transport protocol which
carries applications ranging from bulk data transmission to web browsing. Over
the years, TCP has evolved from original TCP Tahoe~\cite{reno} to the currently
most widely used TCP NewReno~\cite{newreno}. TCP uses Slow Start during startup
phase to probe the capacity of a network path with unknown characteristics. The
TCP probing rate is controlled by its congestion window, \emph{cwnd}, where a
TCP connection can transmit up to \emph{cwnd} amount of unacknowledged packets.

TCP carries 95\% of today's Internet traffic and constitutes 80\% of the total
number of flows in the Internet~\cite{study}. Among those TCP traffic,
short-lived TCP flows spend most of their operational lifetime within the Slow
Start process when \emph{cwnd} ramps up in an exponential manner. Measurement
in~\cite{indication} shows that 85\% of the TCP traffic are short flows. This
implies that the majority of data transmission in the Internet is dominated by
the TCP startup behavior.

In the Slow Start process, \emph{cwnd} is set between one and four TCP packets
initially~\cite{initial1}, and its value is incremented by one packet upon each
reception of an ACK in order to probe and test the available bandwidth. With
this increment, \emph{cwnd} is doubled for each round trip time (RTT) when all
ACKs are returned. As a result, the value of \emph{cwnd} is increased
monotonically with an exponential rate for every RTT until when the network
cannot cope with the amount of transmission from the TCP connection. The
network congestion is signaled by triple duplicated ACKs or more seriously a
Timeout detected by the TCP sender. When this congestion signal is detected,
the TCP connection ends the Slow Start process and the Congestion Avoidance
process takes over the adjustment of \emph{cwnd}. Unlike Slow Start, Congestion
Avoidance maintains a linear increment of \emph{cwnd} every RTT to avoid
congestion.

The exponential increment of $cwnd$ in Slow Start may significantly overshoot
the available bandwidth when probing and testing the bandwidth availability.
This overshooting of $cwnd$ may cause serious congestion and packet loss which
require a long time to recover. To prevent this overshooting, Slow Start
introduces a parameter called Slow Start threshold, \emph{ssthresh}, where when
\emph{cwnd} reaches \emph{ssthresh} where an overshooting is likely, the TCP
connection ends the Slow Start process and lets the Congestion Avoidance
process to take over turning the growth of \emph{cwnd} to a conservative linear
rate.

In general, the current Slow Start process combines an estimation of the
bandwidth availability described by $ssthresh$ and a rate probing algorithm in
order to achieve high bandwidth utilization. Depending on the accuracy of the
bandwidth availability estimation, a corresponding rate probing algorithm can
be designed to achieve a certain high level of bandwidth utilization during the
Slow Start process. For example, in an extreme case where the bandwidth
availability estimation is highly accurate, the rate probing is unnecessary as
a TCP sender can immediately operate at the optimal rate based on the
estimation. In contrast, an inaccurate bandwidth availability estimation should
accompany with a prudent rate probing algorithm to compensate the inaccuracy of
the estimation.

However, Slow Start is known to be extremely inefficient. Two obvious problems
in the current Slow Start algorithm design leads to this inefficiency, and
these problems are particularly severe in long fat networks
(LFNs)~\cite{count1,count2}. The first obvious problem is the blind initial
setting of \emph{ssthresh} due to lack of bandwidth availability estimation.
With the blind initial setting of \emph{ssthresh}, a prudent rate probing
algorithm should be sought. However, the current Slow Start process uses
exponential rate increment probing which is an aggressive rate probing
algorithm. This combination amplifies the performance penalty of the problem.
Precisely, when \emph{ssthresh} is set too high compared to the bandwidth-delay
product (BDP), which represents the capacity of a network pipe, a TCP
connection may inject more packets into a network causing congestion. This
problem is serious in LFNs, because \emph{cwnd} is doubled every RTT, and this
aggressive increase may easily cause burst losses and consequent
Timeout~\cite{reno} by overshooting the network capacity. Conversely, when
\emph{ssthresh} is set too low, a TCP connection will exit Slow Start and enter
Congestion Avoidance prematurely. Thus, \emph{cwnd} may take a long time to
reach an optimal operating point that matches the capacity of the LFN. Both
cases cause low link utilization. Here, we call this drawback the blind
\emph{ssthresh} setting problem.

Another problem in the current Slow Start is related to the temporal queue
buildup occurs during Slow Start. Packets buildup in a buffer occurs when a TCP
connection increases $cwnd$ and transmits more packets within a RTT round. An
adequate buffer size is critical to hold this buildup of packets, otherwise
packet loss will occur. Since Slow Start disregards of the backlog status in
the bottleneck buffer, packet losses may occur even before $cwnd$ has reached
the available bandwidth, which significantly degrades the performance. This
problem is significant when the buffer size of the bottleneck router is much
smaller than the BDP. Here, we call this drawback the temporal queue buildup
problem. This problem has been observed and discussed in \cite{tcpw-a}.

The current efforts in improving the performance of TCP Slow Start largely
focus on improving bandwidth
estimations~\cite{buffer1,buffer2,buffer3,westwood,RCE} for optimal $sshtresh$
setting and/or designing an appropriate rate probing algorithm based on the
reliability of bandwidth availability estimation. However, due to the limited
ability of a TCP sender in observing the network resource, together with the
fast changing network bandwidth availability, the accuracy of bandwidth
availability estimation is largely uncertain giving no basis for the design of
an adequate rate probing algorithm for optimizing the Slow Start performance.

Recognizing the challenges in finding optimal settings for the $sshtresh$ and
the probing algorithm, in this paper, we take a different approach that
bypasses the need for \emph{ssthresh} setting which influences greatly to the
performance. Our solution, called \emph{threshold-less slow start} or
SSthreshless Start (pronounced as s-thresh-less start), is a sender-side
enhancement that offers immediate benefits upon deployment. The key idea of our
method is that it makes use of backlog status at the bottleneck buffer,
monitored by RTT to refine the \emph{cwnd} ramping up behavior and adaptively
adjust probing rate to meet the available network capacity. SSthreshless Start
proposes alternating between an exponential and a linear growth of $cwnd$ based
on backlog status during the TCP startup phase. Our preliminary results
reported in~\cite{Lu2010} have shown encouraging performance gain in TCP
startup. In this paper, extensive simulation and in depth investigations are
conducted to evaluate the benefit of rate alternation on TCP startup
performance.

Briefly, The alternating of two rates achieves benefits in three aspects.
Firstly, it eliminates the need of \emph{ssthresh} to decide when Congestion
Avoidance should take over to end the exponential growth of $cwnd$. Without
\emph{ssthresh}, the blind \emph{ssthresh} setting problem does not exist in
SSthreshless Start. In other words, the network status detection does not
translate to \emph{ssthresh}, instead, the status is used directly to control
the rate probing algorithm between aggressive and prudent modes, where $cwnd$
is increased continuously alternating between an exponential and a linear rates
until a congestion signal is detected. Secondly, SSthreshless Start monitors
the backlog status and switches the growth rate of $cwnd$ to linear when queue
buildup is observed. This prevents continuous queue buildup in the buffer and
hence avoids the temporal queue buildup problem from materialized into a packet
lost event before the available bandwidth is reached. Finally, since $cwnd$
increases monotonically during SSthreshless Start, packet loss is inevitable
due to the finite availability of the bandwidth. However, as network congestion
approaches, the number of backlogged packets at the bottleneck buffer will have
increases. This will signal SSthreshless Start turning to linear growth rate
for $cwnd$. The preemptive switch to a linear growth rate for $cwnd$ as network
congestion approaches allows a fast recovery when a packet loss event
eventually occurs.

We implement our TCP startup solution and combine it with NewReno. NewReno is
chosen because several existing startup modifications, for example Hoe's Change
\cite{hoe} and Limited Slow Start~\cite{limited}, are also refined based on
NewReno. This also allows a direct comparison to existing modifications.
Comparing with traditional Slow Start and existing modifications, our
enhancement shows significant improvement in link utilization during the
startup process with various BDP and buffer configurations. Besides, our
enhancement also shows good convergence behavior without adversely affecting
coexisting TCP connections. Therefore, the throughput gain during startup is
achieved by using the spared bandwidth effectively rather than aggressively
depriving bandwidth from other co-existing TCP connections.

The remainder of the paper is organized as follows. We start by demonstrating
the problems with Slow Start by simulations in the next section. After
summarizing some related works in Section III, we describe the SSthreshless
Start in Section IV, validate it through intensive simulation experiments in
Section V, and we finally conclude this paper in Section VI.


\section{Problems with Traditional Slow Start}

\begin{figure}[t]
\centering
\includegraphics[width=0.48\textwidth]{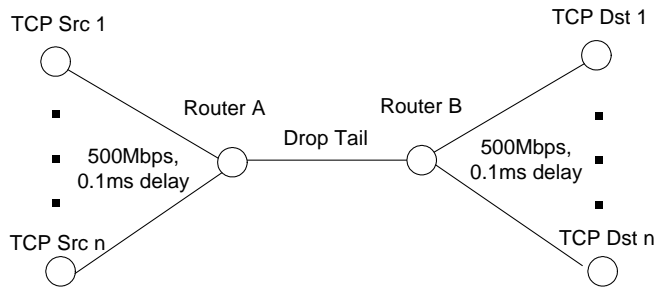}
\caption{Network simulation topology.} \label{fig_topology}
\end{figure}

To illustrate the inefficiency of Slow Start in a LFN, we conduct simulation
experiments using ns-2.34~\cite{NS}. Fig.~\ref{fig_topology} shows our
considered network topology used commonly for this purpose of study. In
Fig~\ref{fig_topology}, TCP Src represents the TCP sender and TCP Dst
represents the TCP receiver. Routers A and B are two droptail bottleneck
routers. Side links are all with bandwidth of $500$ Mbps, and one-way delay of
$0.1$ ms. Between the two routers there is a bottleneck link with $40$ Mbps
bandwidth and $50$ ms one-way delay. For convenience, congestion window size is measured
in number of packets, and the packet size is $1000$ bytes while ACK is set to
$40$ bytes long. This gives the BDP value to be $500$ packets. The bottleneck
router is with $250$ packets buffer size (BDP/2). TCP sources run NewReno with
traditional Slow Start. All the other simulation experiments conducted in this
paper also use this dumbbell topology with varying buffer size, one-way delay
and bandwidth.


In traditional Slow Start~\cite{reno}, before a TCP connection starts, initial
\emph{ssthresh} is set to an arbitrary value, ranging from 4 KB to extremely
high. This blind \emph{ssthresh} setting problem severely degrade TCP startup
performance, especially in LFN. We conduct an simulation to illustrate the
impact of \emph{ssthresh} setting on the performance of Slow Start for short
transfers.

\begin{figure}[t]
 \centering
  \subfigure [Ramping up behavior of NewReno with different \emph{ssthresh} setting] {
    \label{3ssthresh}
    \centering
    \includegraphics[angle=-90,width=0.48\textwidth]{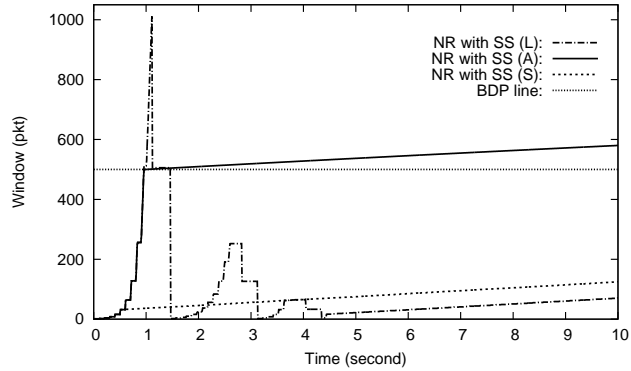}}
  \centering
    \subfigure [Sequence number of packets being sent] {
    \label{seq}
    \centering
    \includegraphics[width=0.5\textwidth]{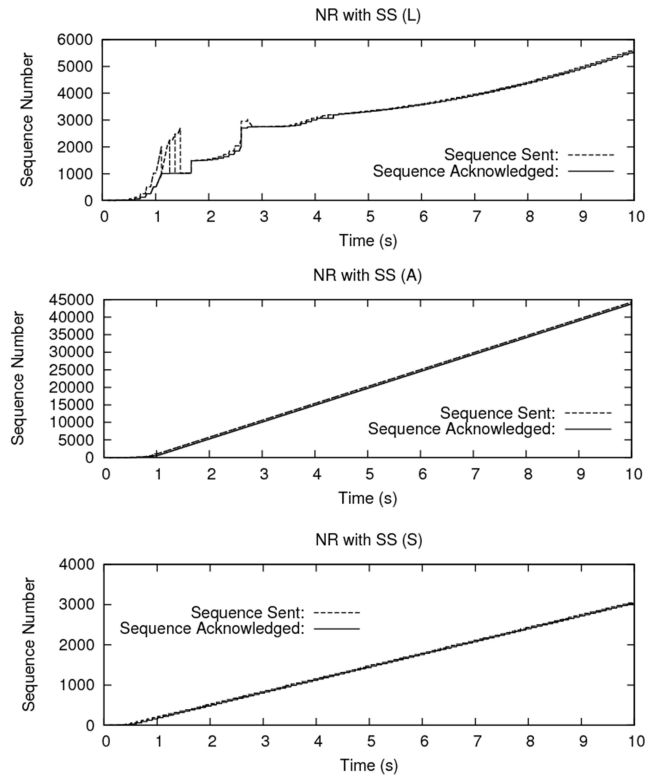}}
  \caption{Comparison of NewReno with different \emph{ssthresh} setting}
\label{fig_single}
\end{figure}

In our setup, we consider a single TCP connection with three cases of different
\emph{ssthresh} values where one is higher than, one is equal to and one is low
than the BDP value. Precisely, we set the \emph{ssthresh} values to 5000, 500
and 32 packets in each simulation and label them as ``NR with SS (L)", ``NR
with SS (A)", ``NR with SS (S)", respectively. We plot the \emph{cwnd} and
sequence number evolution of the three studied cases in Fig.~\ref{fig_single}.

As can been seen from the results in Fig.~\ref{fig_single}(a), ``NR with SS
(L)" that overestimates the BDP quickly overshoots the BDP and produces burst
losses at the router. These burst losses cause a series of Timeout events in
TCP that forces its \emph{cwnd} to exit exponential grow phase prematurely
after Timeout restart. When \emph{ssthresh} is set much lower than the BDP in
the case of ``NR with SS (S)", we see that the TCP connection exits Slow Start
and switches to Congestion Avoidance phase prematurely. Both cases result in
very low bandwidth utilization as illustrated in Fig.~\ref{fig_single}(b).
Whereas if \emph{ssthresh} is set appropriately, we see the best performance
amount all as the TCP connection rapidly grows to the BDP, exits the Slow Start
and maintain stable \emph{cwnd} throughout the simulation period. The overshooting of BDP by
\emph{cwnd} in the case of ``NR with SS (L)" is clearly indicated by the great
discrepancy between the sent and received sequence numbers as shown in
Fig.~\ref{fig_single}(b).

We further record the link utilization and the highest sequence number of the
packet being sent for each case in Table~\ref{table1}. By comparing the
throughput, it is clear that an inadequate setting of \emph{ssthresh} affects
greatly the transmission capability of short connections.


\begin{table}
\centering
\caption{\footnotesize Link utilization and highest packet
sequence of startup schemes  ( first 10s).}
\begin{tabular}{|c|c|c|} \hline
Protocol& Link Utilization & Highest Sequence No. Sent\\ \hline \hline
NR with SS (L) & 11.13\%   &   5636\\ \hline 
NR with SS (A) & 87.50\%   & 44249 \\ \hline
NR with SS (S) & 6.08\%   &   3010\\ \hline
\end{tabular}
\label{table1}
\end{table}

\begin{figure}
\centering
\includegraphics[angle=-90,width=0.48\textwidth]{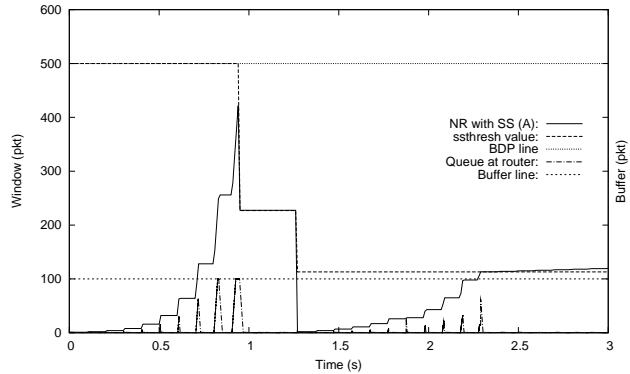}
\caption{Ramping up behavior of NewReno with Slow Start ($ssthresh$=500pkts, Buffer Size=1/5BDP).} \label{3s}
\end{figure}

However, even the \emph{ssthresh} is set to match the BDP at the start, we
found that $cwnd$ may still fail to reach the BDP as also reported by others in
the literature~\cite{short}. To illustrate the effect, we test a TCP connection
whose initial \emph{ssthresh} matches the BDP and the buffer size is set to
only 0.2 times of the BDP. The $cwnd$ and buffer utilization evolutions during
the startup are plotted in Fig.\ref{3s} .

As can be seen, as $cwnd$ reaches over 250 packets, the bottleneck buffer hits
its maximum utilization and a small amount of packet drops is recorded. Since
the TCP sender takes an RTT period to realize the packet loss, it continues to
ramp up its $cwnd$ by doubling the value to over 400 packets resulting
significant packet drops.

Upon detection of the first triple duplicate ACKs, the TCP connection switches
to Fast Recovery and cuts $ssthresh$ to just below 250. The increment of $cwnd$
stalls as only duplicate ACKs are returned. During this period, since $cwnd$
remains constant, the queue barely builds up in the buffer. Being unable to
recover all lost packets, a Timeout event finally occurs at around 1.3s where
$cwnd$ is set to 1 and $ssthresh$ is adjusted to just above 100, and the
connection returns to Slow Start at a much lower $ssthresh$ value. At this
point, the TCP connect has seriously underestimated the available bandwidth
which results in significant underperformance. The cause of this problem is
attributed to the failure of probing rate control when temporal queue buildup
occurs in the bottleneck buffer.

\section{Related Works}\label{sec:literature}

One critical problem of traditional Slow Start performance inefficiency is that
TCP sender lacks the ability to estimate the network condition properly. To
improve TCP startup performance, many approaches have been attempted in the
past to achieve better estimation of network bandwidth availability and/or
design a rate probing mechanism that is less susceptible to the accuracy of the
bandwidth availability estimation. Generally, these efforts to enhance startup
performance can be categorized into four different strategies described below.
\begin{itemize}
    \item The rate probing refinement approach: In this approach, a TCP connection uses
          a different rate probing mechanism than the traditional one to achieve
          better utilization of available network bandwidth. Some proposed
          mechanisms also use dynamic rate probing mechanisms, where returned
          ACKs are used to indicate the network status and adjust the rate probing
          mechanisms.
    \item The bandwidth estimation approach: In this approach, a TCP connection
          performs an estimation of the network to assist rate probing. The estimation may
          perform continuously.
    \item The history-based approach: In this approach, a TCP connection uses
          historical data about the network resource availability cached by
          previous or concurrent connections to estimate the current network status
          and derive optimal parameters for the TCP connection to start.
    \item The router-assisted approach: In this approach, a TCP connection uses
          direct feedbacks from routers to indicate network resources and adjusts
          its sending rate accordingly.
\end{itemize}

\subsection{The Rate Probing Refinement Approach}

The rate probing refinement approach seeks modification of the \emph{cwnd}
ramping up behavior such that the increment of the probing rate is less
susceptible to the accuracy of the initial guess of the bandwidth availability
estimation. A typical example of this approach is the Limited Slow Start
(LSS)~\cite{limited} which uses an additional threshold to prevent the Slow
Start algorithm from increasing too fast. It introduces a new Slow Start
threshold, \emph{max\_ssthresh}, that prevents the probing rate from growing
excessively high. Precisely, when \emph{cwnd} $\leq$ \emph{max\_ssthresh},
\emph{cwnd} doubles for each RTT as in the traditional Slow Start. When
\emph{max\_ssthresh}$<$\emph{cwnd} $\leq$ \emph{ssthresh}, \emph{cwnd} is
increased by a fixed amount of $max\_ssthresh/2$ packets for every RTT. This
condition reduces the growth rate of \emph{cwnd} which in turns reduces the
number of drops during the startup. However, the blind \emph{ssthresh} setting
problem remains unsolved with this approach. Other schemes based on a similar
strategy such as CapStart~\cite{capstart} and Smooth-Start~\cite{smoothstart1,
smoothstart2} suffers the same shortcoming.

In~\cite{vegas}, TCP Vegas has demonstrated that the packet delay at the
bottleneck router can be estimated by observing the RTT of each packet
transmission. This provides a better guidance for a TCP sender to either refine
its rate probing strategy or adjust Slow Start parameters such as
\emph{ssthresh} to enhance its Slow Start performance. Based on the observed
packet delay, TCP Vegas uses a different rate probing strategy, namely, it
doubles \emph{cwnd} every other RTT, and exits the Slow Start phase when the
estimation of packet delay exceeds a certain threshold. This method, however,
often leads to low bandwidth utilization due to premature exiting of Slow Start
as a result of temporary queue buildup in the buffer caused by bursty TCP
transmission~\cite{tcpw-a}. Enhancing the usage of RTT information, Delay-base Slow Start
(DBSS)~\cite{delaybased} uses RTT information to adjust \emph{max\_ssthresh}
which indirectly prevents \emph{cwnd} from overshooting and avoid premature
exiting of the Slow Start phase. However, it requires a threshold on RTT to
function and setting of an appropriate threshold remains a challenge. 

\subsection{The Bandwidth Estimation Approach}

The bandwidth estimation approach aims to solve arbitrary \emph{ssthresh}
setting problem by setting it to some estimated BDP value to mitigate the effect of
overshooting while maintaining the original rate probing strategy. Bandwidth
estimation is first introduced in~\cite{first} by using packet pair bandwidth
measurement technique. Packet pair measurement uses the inter-arrival time
between the ACK pair received at the source to infer the bottleneck bandwidth
along the path. Based on this technique, Hoe~\cite{hoe} proposed to set initial
\emph{ssthresh} to the product of the measured delay and the estimated
bandwidth. However, attribute to the aggressive \emph{cwnd} increase manner,
Hoe's Change may suffer temporary queue overflow and multiple losses when the
bottleneck buffer is not large enough compared to the BDP, or many flows are
coexisting~\cite{tcpw-a}. Several improvements based packet pair
bandwidth measurement have been proposed to enhance Hoe's
method~\cite{Aron,swiftstart1,swiftstart2,packetpair1,pacing}. Nevertheless,
evidenced in~\cite{antihoe1,antihoe2}, the packet pairs technique gives a
reliable estimation of the bottleneck link capacity rather than an available
bandwidth on a network path. Hence, only limited performance gain can be
achieved.

Beside the packet pairs technique, the packet trains estimation measurement
appears to be more reliable for the estimation of the instantaneous available
bandwidth of a path. Early Slow Start Exit (ESSE) proposed in~\cite{pipesize}
uses observation from a series of ACK returning times to estimate the
instantaneous available bandwidth and set the initial \emph{ssthresh} value.
Paced Start proposed in \cite{active} further uses the difference between data
packet train dispersion and ACK train dispersion to interactively for bandwidth
estimation and \emph{ssthresh} setting. TCP Westwood uses Eligible Rate
Estimation (ERE)~\cite{westwood} that relies on ACK train from receiver for
bandwidth estimation. Adaptive
Start~\cite{wang} proposes using ERE to assist the Slow Start.

However, researches carried in~\cite{dispersion1,dispersion2} have shown that
the dispersion of long packet trains does not measure the available bandwidth
in a path, rather, it tells another bandwidth metric known as Average
Dispersion Rate (ADR), the value which is in between an available bandwidth and
a capacity of the path. The direct use of the dispersion of long packet trains
for available bandwidth measure may cause misleading estimation leading to
undesirable performance. Inspired by these findings, Hybrid Slow
Start~\cite{Hybrid} combines ACK train estimation and increase in packet delays in
the the Slow Start algorithm for performance enhancement.

In summary, while these estimation techniques achieve a certain performance
gain compared to Slow Start that uses arbitrary default \emph{ssthresh} value,
the performance gain is limited due to their accuracy in the estimation caused
by various factors. One obvious factor is due to the additional manipulation of
ACK replies in modern TCP operations, such as ACK clustering and
compression~\cite{CA1,compassion}, Delayed ACK~\cite{DACK} are affecting the
accuracy of bandwidth estimation. Other factors such as TCP coarse-grained
clocks~\cite{clock2}, rerouting~\cite{rerouting} and route asymmetry
between forward and reverse path also pose challenges to the accuracy of
bandwidth estimation. Besides, even an accurate bandwidth estimation technique
is achieve, this approach of using bandwidth estimation do not deal with and
hence cannot resolve the temporal queue buildup problem.



\subsection{The History-Based Approach}

The history-based approach makes use of history information cached by previous
or concurrent connections to improve Slow Start performance. It is based on the
assumption that any hosts in the sub-domain would experience similar
performance to distant hosts. Usually schemes fall into this catalog is
intended for a restart transmission on connections that have been idle for a
long time to benefit some certain applications (i.e., web browsing).


Transaction TCP~\cite{transactions1, transactions2} caches previous connection
count history in order to save the three-way handshake in certain situations to
speed up future connection establishment. Expanding on the available historical
information, TCP Control Block Sharing~\cite{interdependence} and Congestion
Manager~\cite{manager} propose sharing of Slow Start related information among
recent or concurrent TCP connections with the same end nodes. Other incremental
enhancements falling within this approach include TCP with Shared Passive
Network Discovery (SPAND) or TCP/SPAND~\cite{spand}~\cite{optimizating1} and Adaptive TCP Slow
Start~\cite{adaptivetcp}.


In summary, the historical-based approach makes use of historical information
to help a new TCP connection tune to a more appropriate sending rate. However, the
usefulness of the historical information may vanish quickly due to the fast
changing load conditions in the network. Besides, this approach is unable to
benefit TCP startup performance when historical information does not exist. For
example, when a connection is established to a new destination, traditional
Slow Start is adopted instead.

\subsection{The Router-Assisted Approach}

It is illustrated that assistance from routers for TCP rate control is
effective to achieve high utilization of network bandwidth~\cite{Ariba2009}.
Measured directly at the routers, it offers accurate bandwidth availability
utilization, and the role of rate probing algorithm can be significantly
reduced. Quick-Start~\cite{quickstart1} and XCP~\cite{XCP} are some typical
examples for this approach. In Quick-Start, a TCP sender advertises a desired
sending rate during the three-way handshake to let the network (each hop along
the path) approve, reject or reduce the requested sending rate. This way, a
sender can quickly tune to an appropriate rate without the time consuming
probing procedure. Comparatively, XCP proposes a more fine-grained feedback to
TCP senders for them to decide their sending rates.

In summary, while the router-assisted approach gives potential to significantly
improve the utilization of networks especially during the startup phase of a
TCP connection, they require special operations in routers which prevents them
from immediate deployment and thus their attractiveness is not high.

\section{The Enhancement}

We shall introduce a novel startup scheme, called SSthreshless Start, with the
goal to address the two aforementioned problems in the traditional Slow Start.
As discussed in Section~\ref{sec:literature}, with the accuracy limitation in
the bandwidth estimation and the history-based approaches, and the deployment
drawback in the router-assisted approach, we argue that the rate probing
refinement approach remains a potential approach that can offer significant
performance gain in Slow Start with immediate deployment. However, the main
challenge of the rate probing refinement approach is the ability to quickly
probe available bandwidth for the setting of the sending rate to ensure high
utilization based on a certain bandwidth availability estimation translated
into \emph{ssthresh} setting. Recognizing the challenges in finding an optimal
setting for \emph{ssthresh} based on a certain bandwidth availability
estimation and an adequate rate probing algorithm, we take a difference that
bypasses the need for \emph{ssthresh}. With this novel attempt, we design a new
startup scheme that not only achieves efficient sending rate, but also copes
well with the temporal queue buildup problem. Owing to needless of \emph{ssthresh}, we
call our startup scheme \emph{threshold-less slow start} or SSthreshless Start.
We detail SSthreshless Start in the following subsections.


\subsection{Backlogged Packet Detecting}

TCP Vegas is known as a delay-based congestion control mechanism since it uses
RTT for each packet transmission to estimate the backlog status of the buffer
to adjust its congestion control strategy. Past research~\cite{vegas1,vegas}
has shown that this estimation, in terms of the delayed packets due to
buffering at the bottleneck router, leads to a more accurate estimation of
network traffic load condition. Capitalizing on this effective estimation, we
reuse this estimation mechanism in our proposed SSthreshless Start.

In TCP Vegas, the throughput difference is calculated by
\begin{displaymath}
Diff=(Expected-Actual)=\left(\frac{cwnd}{BaseRTT}-\frac{cwnd}{RTT}\right)
\end{displaymath}
where \emph{BaseRTT} is the minimum of all measured RTT, and \emph{RTT} is the
actual round trip time of a tagged packet. Denote the delayed packets at
bottleneck buffer by $N$, we have,
\begin{displaymath}
RTT=BaseRTT+N/Actual.
\end{displaymath}

Rearranging the above equation, we obtain
\begin{equation} \label{eqn:n}
N = \left(\frac{cwnd}{BaseRTT}-\frac{cwnd}{RTT}\right)\times BaseRTT.
\end{equation}


During startup phase we can use \eqref{eqn:n} to calculate the delayed packets
at bottleneck buffer. This provides the information of backlog status for our
SSthreshless Start.

\subsection{SSthreshless Start}

The key idea of SSthreshless Start is that it makes use of backlog status at
the bottleneck buffer, monitored by RTT to refine the \emph{cwnd} ramping up
behavior and adaptively adjust probing rate to meet the available network
capacity. Rather than translating the network status into $ssthresh$, the
network status is directly used to control the rate probing algorithm.

We propose a two-mode operation in the rate probing procedure, namely, Linear
Increase Mode and Adjustive Increase Mode, each mode is intended to operate in
the situation when the queue buildup is detected or not detected,
respectively. Recall that the estimated delayed packets number is $N$, a
certain number of estimated delayed packets, $N \geq \beta$, can be used to
signal a packet building up event at the bottleneck router. The quantity
$\beta$ is a design time protocol parameter for SSthreshless Start to switch
between Linear Increase and Adjustive Increase modes.
While $\beta$ is arbitrary set, we shall show that the Slow Start performance
is insensitive to this threshold.

Based on \eqref{eqn:n}, SSthreshless Start measures $N$, the estimation of
backlog at the bottleneck router, and compares with the threshold $\beta$. If
the estimated number of backlog packets exceeds $\beta$, we assume that the
bottleneck router is experiencing packet building up. Once the backlog is clear
below $\beta$, the TCP sender is said to have experienced one congestive event.
In our scheme, the TCP sender monitors the congestive status and records the
total number of congestive event experienced for rate probing purpose.

Each TCP connection begins with the binary exponential increase of \emph{cwnd}
as in the traditional Slow Start. Different from the traditional Slow Start,
each TCP continuously monitors the backlog packets. When the monitored number
of the backlog packets exceeds $\beta$ indicating the occurring of packet
building up, SSthreshless Start takes over the control of \emph{cwnd}.

SSthreshless Start operates in either Linear Increase Mode or Adjustive
Increase Mode, and it always starts in Linear Increase Mode. In Linear Increase
Mode, the TCP connection increases its \emph{cwnd} by one for every RTT which
confines the \emph{cwnd} increment to a linear rate. SSthreshless Start is
activated when the estimated backlog packets exceeds $\beta$ for the first time, thus
starting SSthreshless Start in this conservative increase manner helps clear
temporary buildup by slowing down the transmission from the source to the
buffer to avoid buffer overflow and multiple losses. SSthreshless Start remains
in this mode as long as the monitored number of backlog packets exceeds
$\beta$.

Once the monitored number of backlog packets falls below $\beta$, SSthreshless
Start switches to Adjustive Increase Mode where \emph{cwnd} increment rate
turns aggressive again. The aggressiveness of the increment also depends
on the number of encounters of congestive events. In our design, the more
congestive events a TCP connection encounters, the milder is the increment of
its \emph{cwnd}. This is because as \emph{cwnd} increases monotonically in
either Linear Increase Mode or Adjustive Increase Mode, the likelihood of
\emph{cwnd} reaching the available bandwidth also increases, and a milder
increment in \emph{cwnd} should be used to prevent it from overshooting the
available bandwidth causing serious losses.

Note that, different from Delay-based TCP startup schemes (i.e., TCP Vegas and
DBSS), where delay-based information is used the find the threshold of exiting
startup phase, SSthreshless Start uses backlog status to dynamically switch
\emph{cwnd} ramping up behavior in an exponential-linear cycles, adaptively
seizing the available bandwidth. SSthreshless Start exits when packet losses
occur. The pseudo code of SSthreshless Start is given in the following.

\begin{algorithm}
\caption{SSthreshless Start}
\label{alg1}
\begin{algorithmic}
\IF {(three DUPACKs are received)/*startup phase exits*/}
    \STATE \emph{ssthresh=cwnd/2};
    \STATE \emph{Congestion\_Event\_No=0};
    \STATE \emph{Congestive\_Status=0 // congestive status of last RTT};
    \STATE \emph{/*switch to Congestion Avoidance Mode*/}
\ELSE
    \IF {($N\geq\beta$)}
         \STATE \emph{Congestive\_Status=1};
         \STATE  $cwnd+=\frac{1}{cwnd}$; \emph{on each ACK}
         \STATE \emph{/*Linear Increase Mode*/}
    \ELSE
         \IF {(\emph{Congestive\_Status}=1)}
         \STATE \emph{Congestion\_Event\_No++};
         \STATE \emph{Congestive\_Status=0};
         \ENDIF
         \STATE $cwnd+=\max\left(\frac{1}{cwnd}, \frac{1}{2^{Congestion\_Event\_No}}\right)$;
         \STATE \emph{for every ACK}
         \STATE \emph{/*Adjustive Increase Mode*/}
    \ENDIF
\ENDIF

\end{algorithmic}
\end{algorithm}

In the above pseudo code, \emph{Congestive\_Event\_No} indicates the times of
congestive events occurred with its initial value set to $0$. According to our
design shown in Algorithm \ref{alg1}, the increment of \emph{cwnd} in Adjustive
Increase Mode is set between $\frac{1}{cwnd}$ and 1 for every ACK. In other
words, for every RTT, $cwnd$ is increased by a value between 1 and $cwnd$.

\section{Performance evaluation}

In this section, we present numerical results of SSthreshless Start, compared
with the tradition Slow Start and other variants, given different network
environments with dissimilar parameter settings. We first evaluate the parameter
setting of SSthreshless Start, then demonstrate the ramping up behavior and
throughput advantages over other variants. Finally, we also show the fairness
and friendliness of our SSthreshless Start.

~\subsection{Parameter Setting}

\begin{figure}
\centering
\includegraphics[angle=-90,width=0.48\textwidth]{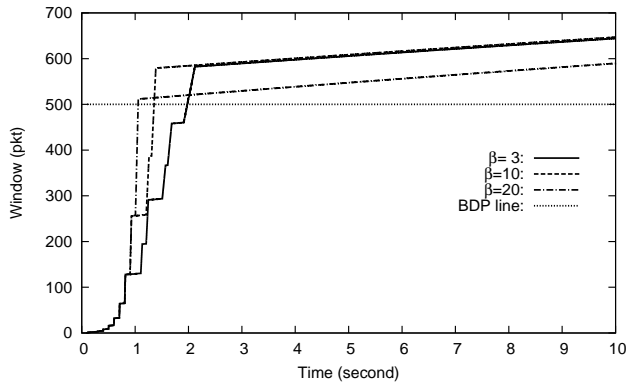}
\caption{SSthreshless Start \emph{cwnd} evolution under different values of
$\beta$.} \label{fig_Parameter}
\end{figure}

\begin{table}
\centering
\caption{\footnotesize Performance comparison of SSthreshless Start under different value of $\beta$  (first 10s).} \begin{tabular}{|c|c|c|} \hline
$\beta$ & Link Utilization & Highest Sequence No. Sent \\ \hline \hline
$\beta= 3$ & 82.15\% & 41075 \\ \hline 
$\beta=10$ & 83.95\% & 41973   \\ \hline
$\beta=20$ & 85.39\% & 42699  \\ \hline
\end{tabular}
\label{tableparameter}
\end{table}

In Fig.~\ref{fig_Parameter} we vary the value of switch, $\beta$, to assess the
its sensitivity to the performance of SSthreshless Start. Surprisingly, varying
$\beta$ does not cause much difference in the performance. We present in
Table~\ref{tableparameter} the numerical details for the link utilization and
highest sequence number of packets being sent for three different $\beta$
values from small to large. As can be seen from Table~\ref{tableparameter}, the
performance difference is very small. This indicates that the value of $\beta$
is not a mainly decisive factor in the performance, which makes SSthreshless
Start tolerable to the inaccuracy of TCP
timers~\cite{timer1}~\cite{timer2}, which affects the backlog
estimation, and the setting of $\beta$.

As shown in Fig.~\ref{fig_Parameter}, as $\beta$ goes large, the ramping up
behavior is slightly more aggressive. To an extreme when $\beta$ is set to
infinity, SSthreshless Start will behave like the traditional Slow Start since
the congestive event can never occur. In contrast, when $\beta$ is set to a
small value, the occurrence of the congestive event increases and \emph{cwnd}
grows in a more conservative rate. Based on the past
experiences~\cite{limited,smoothstart1}, conservative growth in \emph{cwnd} may
significantly reduce burst losses, we thus suggest a smaller $\beta$ setting.
We recommend the setting of $\beta=3$ as this setting gives conservative growth
in \emph{cwnd} yet maintain a high link utilization as reported in
Table~\ref{tableparameter}.

~\subsection{SSthreshless Start Ramping up behavior}

\begin{figure}[t]
 \centering
  \subfigure [Ramping up behaviors of NR with SLS] {
    \label{200buffer}
    \centering
    \includegraphics[angle=-90,width=0.50\textwidth]{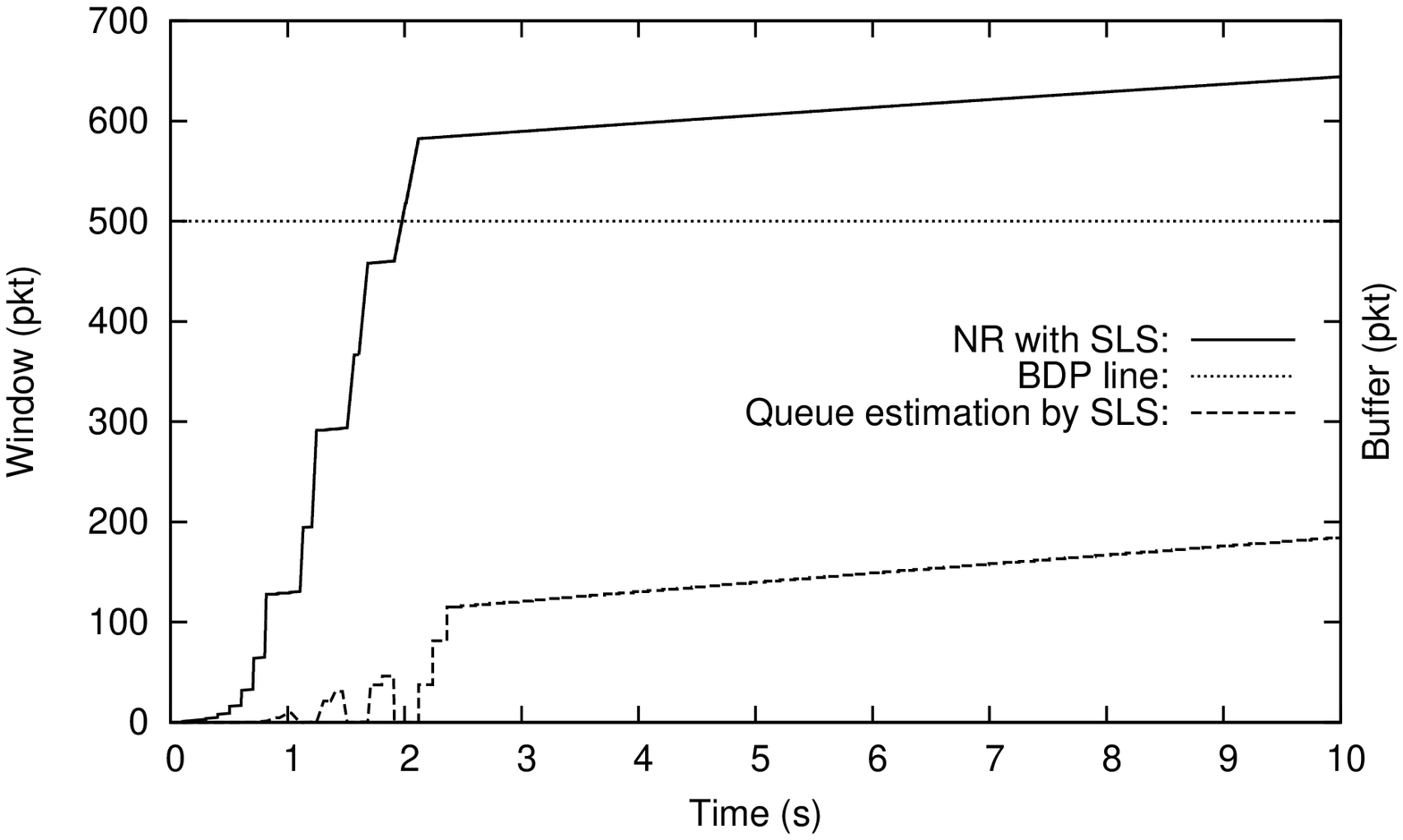}}
  \centering
    \subfigure [Ramping up behavior of NR with SS (A) and TCP Vegas] {
    \label{queue}
    \centering
    \includegraphics[angle=-90,width=0.475\textwidth]{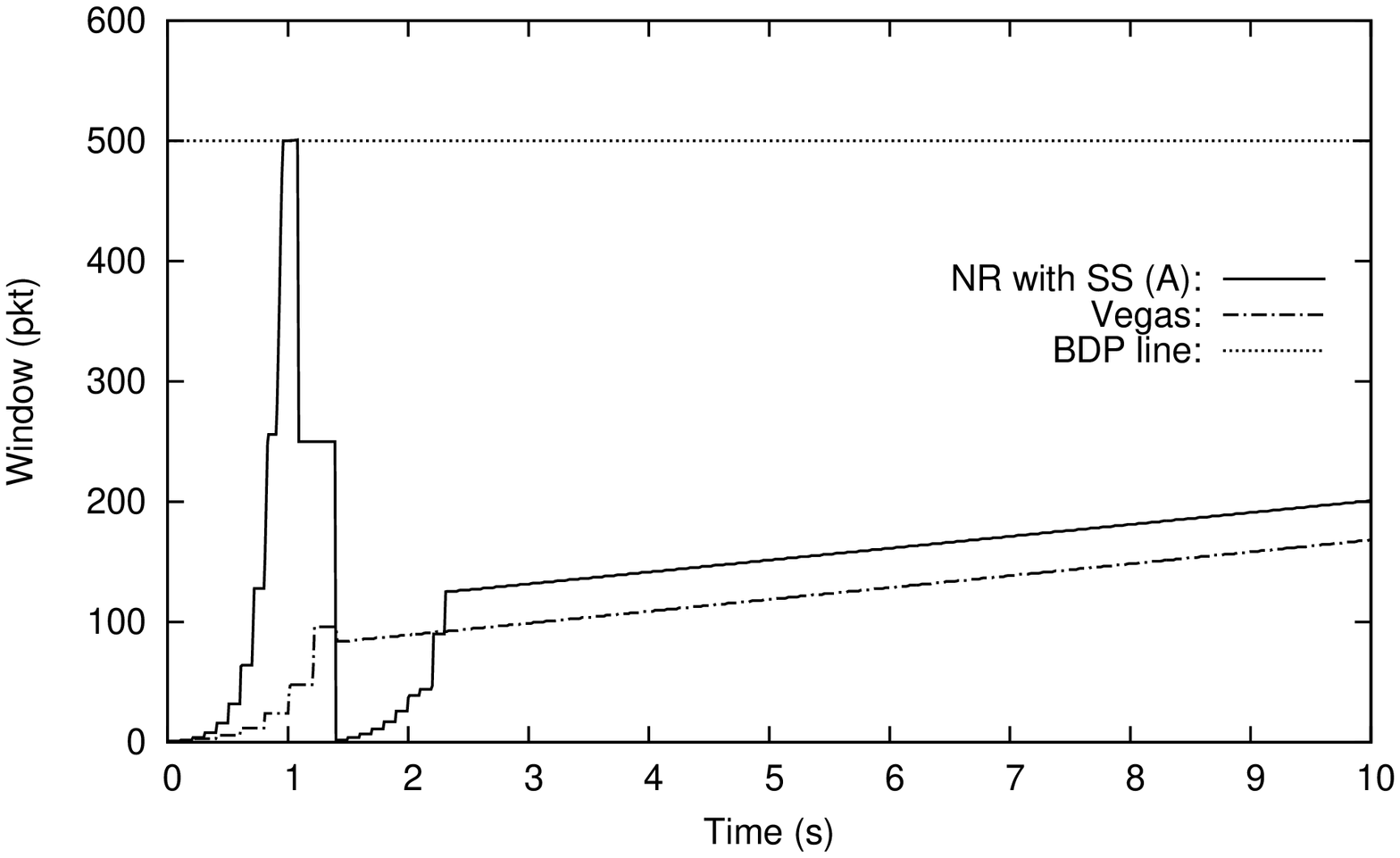}}
  \caption{Comparison of different TCP startup schemes (Buffer=2/5BDP, first 10s)}
\label{three}
\end{figure}

Fig.~\ref{three} compares the single flow ramping up behaviors of SSthreshless
Start, Slow Start and Vegas, along with the monitored backlog at bottleneck
router. The buffer size is set to a small value of 200 packets, or
$\frac{2}{5}$ of the BDP, and \emph{ssthresh} of Slow Start is set to accurate
value of BDP in this case.

Depending on the instantaneous backlog status, SSthreshless Start switches
between exponential and linear rate to increase \emph{cwnd}. It allows
\emph{cwnd} to adaptively ramp up to the BDP in a timely manner. The
\emph{cwnd} value reaches an eligible window size in around 2s, and maintains
high link utilization ever since. By monitoring the backlog status, SSthreshless
Start would have switched to a linear rate just before the occurrence of a
packet loss that ends the SSthreshless Start operation, and this helps the
Congestion Avoidance operation which takes over SSthreshless Start to cope with
the loss.

In comparison, the aggressive Slow Start increases its \emph{cwnd} to the BDP
size in around 1s. However, the temporary queue buildup occurs with packet
drops and following Fast Recovery~\cite{newreno} fails to recover the multiple
losses since Slow Start remains in exponential growth when losses occurred, and
this causes a Timeout. Consequently, the $sshtresh$ is repeatedly adjusted
downward to eventually a small value which fouces the TCP connection to enter Congestion Avoidance phase
prematurely. For Vegas, it also enters Congestion Avoidance prematurely right
after backlog exceeds its threshold. Being too conservative, it manages to
avoid multiple losses but operates in a very low throughput. Over the
evaluation time shown in Fig.~\ref{three}, neither Slow Start nor Vegas is
capable of seizing the eligible BDP where majority of available bandwidth is
left unused. As listed in Table~\ref{threetable}, SSthreshless Start captures
as high as $80\%$ of the link utilization, where Slow Start and Vegas utilize
below $30\%$.

\begin{table}[h]
\centering
\caption{\footnotesize Performance comparison of TCP startup schemes (Buffer=2/5BDP, first 10s).} \begin{tabular}{|c|c|c|} \hline
Scheme & Link Utilization & Highest Sequence No. Sent\\ \hline \hline
NR with SLS & 80.30\%   & 40025 \\ \hline 
NR with SS (A) & 28.20\%  & 14100   \\ \hline
Vegas & 22.35\%   & 11173 \\ \hline
\end{tabular}
\label{threetable}
\end{table}

\subsection{Throughput Comparison}

In this subsection, we compare NewReno with SSthreshless Start (SLS), Hoe's
Change (HC), Limited Slow Start (LSS), Slow Start with small \emph{ssthresh}, $32$ packets (SS (S)),
Slow Start with large \emph{ssthresh}, 5000 packets, (SS (L)), and TCP
Vegas. It shows that the SSthreshless Start significantly improves startup
performance with regards to various buffer size, one-way delay, and bandwidth.
To focus on the startup performance, we only evaluate the performance in the
first $20$ seconds. In addition, we use Throughput Ratio, calculated as the
throughput of SSthreshless Start over other variants, as a measure to evaluate
the performance enhancement of our proposal.

\begin{figure}
\centering
\includegraphics[angle=-90, width=0.48\textwidth]{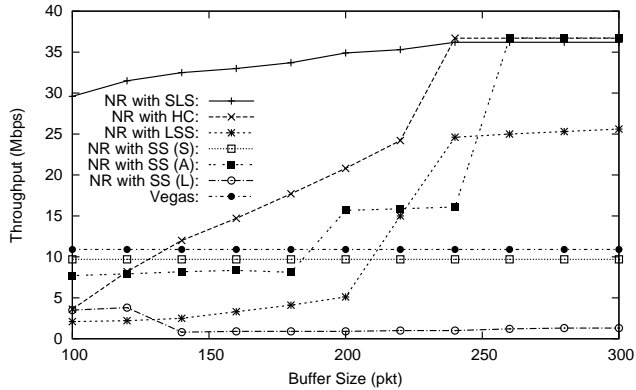}
\caption{NewReno (NR) throughput versus buffer size (first $20$s).} \label{fig_Buffer}
\end{figure}

In Fig.~\ref{fig_Buffer}, we fix the bandwidth to $40$~Mbps, delay to $50$~ms,
and vary the buffer size from $100$ packets to $300$ packets to study impact of
buffer sizes on the performance. In this case, we also evaluate the Slow Start
with \emph{ssthresh} setting to the accurate BDP
size, 500 (SS (A)), to show the buffer robustness of our proposal. It is
evident that high throughput is achieved by our SSthreshless Start in all the
test cases. Also as can be seen, when the buffer size is small, Hoe's Change,
Limited Slow Start and even Slow Start with \emph{ssthresh} set to accurate BDP
suffer severe performance degradation. These startup algorithms fail to obtain
a high throughput even with the help of ample buffer size. The case with a
small buffer size, $1/5$ BDP, is shown in Table~\ref{table4}. The performance
benefit of SSthreshless Start is significant, which gains up to 3 to 14 times
over other variants.

\begin{table}[h]
\centering
\caption{\footnotesize  Performance comparison of TCP startup schemes (40Mbps bottleneck bandwidth, 50ms one-way bottleneck delay, Buffer Size=1/5BDP ).}
\begin{tabular}{|c|c|c|} \hline
Scheme & Link Utilization & Throughput Ratio\\ \hline \hline
NR with SLS & 74.08 \%  &   100.00\% \\ \hline 
NR with HC & 9.05\%  &   820.77\%  \\ \hline
NR with LSS & 5.33\%  &   1391.08\% \\ \hline
NR with SS (S) &   24.25\% & 305.46\% \\ \hline
NR with SS (A) &   19.02\% & 305.46\% \\ \hline
NR with SS (L) & 4.55\%  &   849.00\% \\ \hline
Vegas & 10.11\%  &   271.34\% \\ \hline
\end{tabular}
\label{table4}
\end{table}

\begin{figure}
\centering
\includegraphics[angle=-90,width=0.48\textwidth]{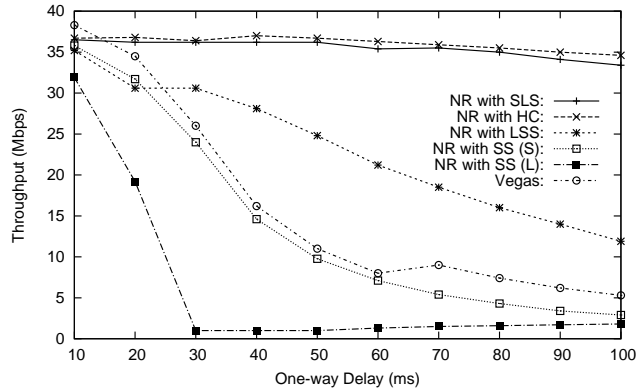}
\caption{NewReno (NR) throughput versus delay (first $20$ s).}
\label{fig_Delay}
\end{figure}

As aforementioned, one of characteristic of LFN is long link delay. Thus, to
assess performance with long RTT, we very the bottleneck one-way delay from
$10$ ms to $100$ ms. The bandwidth and buffer size are fixed to $40$ Mbps and
BDP/2, respectively. Fig.~\ref{fig_Delay} shows the throughput comparison under
this scenario. The subtle changes in the throughput of NewReno with
SSthreshless Start and Hoe's Change shows their ability to scale well with long
delay, while other startup algorithms suffer from performance degradation as
delay increases. The case with largest tested one-way delay, 100~ms, is listed
in Table~\ref{table3}. Note that the throughput of SSthreshless Start achieves
over 18 times that of the traditional Slow Start.

\begin{table}[h]
\centering
\label{table3}\caption{\footnotesize Performance comparison of TCP startup schemes  (40Mbps bottleneck bandwidth, 100ms one-way bottleneck delay, buffer size=1/2BDP).} \begin{tabular}{|c|c|c|} \hline
Scheme & Link Utilization & Throughput Ratio\\ \hline \hline
NR with SLS & 83.35\%  &   100.00\% \\ \hline 
NR with HC & 86.40\%  &   96.70\%  \\ \hline
NR with LSS & 29.80\%  &   280.40\%  \\ \hline
NR with SS (S) &   6.55\% & 459.07\%  \\ \hline
NR with SS (L) & 4.55\%  &   1836.26\% \\ \hline
Vegas  & 10.11\%  &   623.95\% \\ \hline
\end{tabular}
\end{table}

\begin{figure}
\centering
\includegraphics[angle=-90,width=0.48\textwidth]{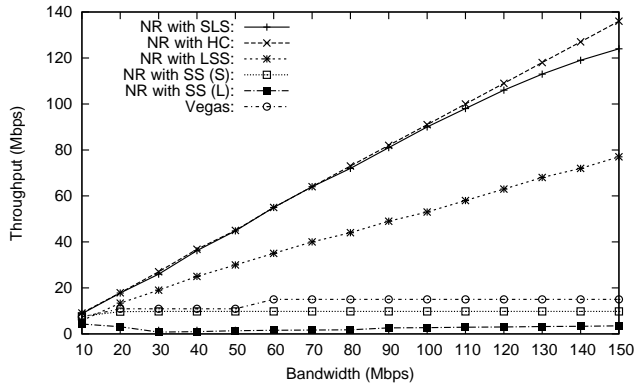}
\caption{NewReno (NR) throughput versus bottleneck bandwidth (first $20$ s).} \label{fig_Bandwidth}
\end{figure}

To evaluate the performance of SSthreshless Start with respect to another
characteristic of LFN, high bandwidth, we vary the bottleneck bandwidth from
$10$ Mbps to $150$ Mbps. Besides, we fix the bottleneck one-way delay to
$50$~ms and buffer size to BDP/2. Fig.~\ref{fig_Bandwidth} reports on NewReno
throughput achievements with different startup schemes under this scenario. It
is shown that, NewReno with SSthreshless Start and Hoe's Change scale well with
a wide range of bandwidth. Other schemes lack the ability to adapt to network
bandwidth effectively, leading to poor throughput achievements. The case with
largest tested bandwidth, 150Mbps, is listed in Table~\ref{table2}. Notably,
SSthreshless Start outperforms the traditional Slow Start just over 35 times in
terms of throughput.

\subsection{Dynamic Bandwidth}
\begin{figure}
 \centering
  \subfigure [Ramping up behavior of NewReno with SLS] {
    \label{udpcwnd}
    \centering
    \includegraphics[angle=-90,width=0.50\textwidth]{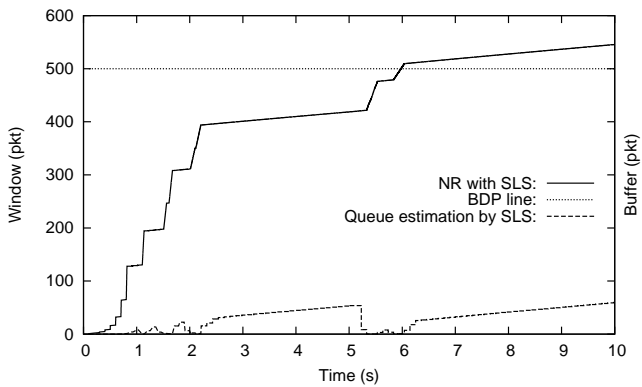}}
  \centering
    \subfigure [Ramping up behavior of NR with SS (A) and TCP Vegas] {
    \label{udpqueue}
    \centering
    \includegraphics[angle=-90,width=0.475\textwidth]{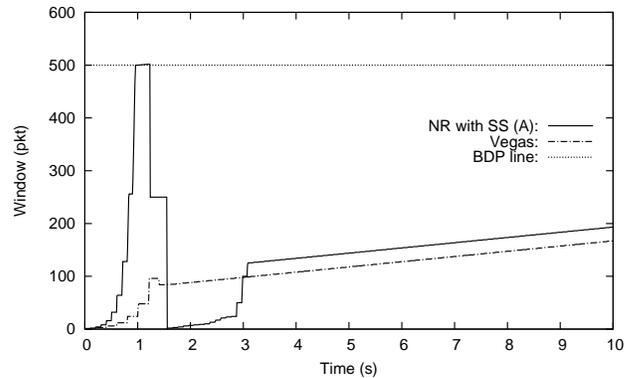}}
  \caption{Comparison of different TCP startup schemes under UDP cross-traffic (10Mbps UDP traffic starts at 1 sec and stops at 5sec)}
\label{fig_UDP}
\end{figure}

Considering that available bandwidth may change several times during the
startup phase of a TCP session under a dynamic environment (i.e., other
connections may join or leave the link), a well-performed startup scheme should
be aware of the instantaneous available bandwidth to adjust the \emph{cwnd}
ramping up strategy. To assess the capability of SSthreshless Start in the
network with dynamic network load, we add a burst UDP cross-traffic set to
10 Mbps, starting at the first second and stopping at the fifth second.
Fig.~\ref{fig_UDP} shows the comparison of SSthreshless Start and Slow Start
under this scenario.


During the first second, $cwnd$ of SSthreshless Start ramps up just as usual.
After initiating the burst of UDP traffic, SSthreshless Start detects the
decrease of available bandwidth quickly through the backlogged queue, and
accordingly, halves \emph{cwnd} growth rate each time when congestive event
happens. After reaching the available bandwidth, \emph{cwnd} turns linear
increment. Then, right after the termination of UDP traffic flow, SSthreshless
Start detects the clearing up of bottleneck backlog and alternates \emph{cwnd}
growth rate back to exponential again. This simulation shows that the
alternation between exponential and linear growth rate of \emph{cwnd} can cope
well with dynamic changing bandwidth availability.

\begin{table}[h]
\centering
\label{table2}\caption{\footnotesize Performance comparison of TCP startup schemes (50ms one-way delay, 150Mbps bottleneck bandwidth, buffer size=1/2BDP).} \begin{tabular}{|c|c|c|} \hline
Scheme & Link Utilization & Throughput Ratio\\ \hline \hline
NR with SLS & 82.84\%  &   100.00\% \\ \hline
NR with HC & 90.58\%  &   91.45\%  \\ \hline
NR with LSS & 51.54\%  &   160.71\%  \\ \hline
NR with SS (S) &   6.55\% & 1265.38\%  \\ \hline
NR with SS (L) & 2.37\%  &   3500.28\% \\ \hline
Vegas  & 10.11\%  &   819.66\%  \\ \hline
\end{tabular}
\end{table}

On the other hand, due to the aggressive and blind increase strategy, Slow
Start incurs losses upon the presence of UDP flow. Following Fast Recovery
fails to recover the multiple losses which leads to a consequent Timeout.

In addition, to compare the transmission capability, the Link Utilization and
the highest sequence number of packets being sent are recorded in Table~\ref{table5}. The amount of
data SSthreshless Start manages to send is almost two times larger than that of
Slow Start. This makes SSthreshless Start greatly benefit short flows, that
lasts only for several seconds under a dynamic environment.

\subsection{Friendliness to Slow Start}

\begin{figure}
\centering
\includegraphics[angle=-90, width=0.48\textwidth]{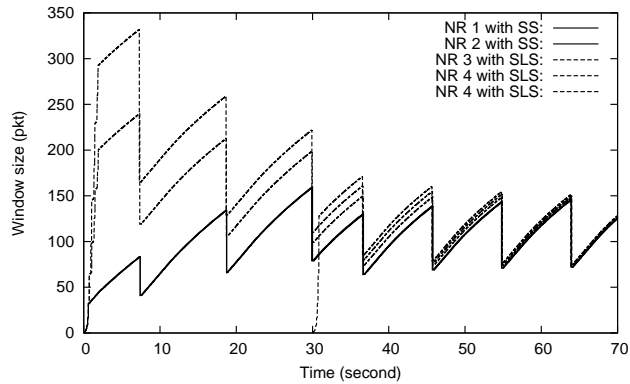}
\caption{Co-existence of multiple SSthreshless Start and Slow Start
connections. (\emph{cwnd} of two NewReno flows overlap)} \label{fig_five-flow}
\end{figure}

Fig.~\ref{fig_five-flow} shows the coexistence of multiple SSthreshless Start
and Slow Start connections. We consider five NewReno connections, in which
connections $1$ and $2$ are NewReno with Slow Start (\emph{ssthresh} = 32
packets) and connections $3$, $4$, $5$ are NewReno with SSthreshless Start.
Connections $1$, $2$, $3$, $4$ start at $0$ second to investigate the effect of
SSthreshless Start and Slow Start startup at the same time. Connection $5$
starts at $30$th second to estimate the effect of SSthreshless Start on
existing TCP connections. It is shown that SSthreshless Start utilizes network
bandwidth left unused by Slow Start connections at the very beginning. After
several rounds of synchronized packet losses, $cwnd$ for each connection
converges to a same value. When connection $5$ joins the network, it does not
adversely affect existing TCP connections. It is able to probing rate quickly to
reach the state of other concurrent TCP connections. Finally, all five
connections converge to the same window size, which is around $100$ packet
sizes or one-fifth of the BDP. Each connection utilizes its own share fairly,
demonstrating the friendliness of NewReno with SSthreshless Start in bandwidth
sharing with the traditional Slow Start.

\begin{table}[h]
\centering
\label{table5}\caption{\footnotesize Performance comparison of NewReno with SSthreshless Start and Slow Start under UDP cross-traffic ( first 10s).}
\begin{tabular}{|c|c|c|} \hline
Scheme & Throughput (Mbps) & Highest Sequence No. Sent\\ \hline \hline
NR with SLS & 27.73   &  34663 \\ \hline 
NR with SS (S)  & 10.50  & 13126  \\ \hline
Vegas  & 8.86  & 11080  \\ \hline \end{tabular}
\end{table}

\section{Conclusions}

In this paper, we present a novel sender-side enhancement, SSthreshless
Start, to improve TCP startup performance in long fat networks. The key idea is
to make use of backlog status at the bottleneck buffer, monitored by RTT to
refine the \emph{cwnd} ramping up behavior, dynamically and adaptively
adjusting probing rate to reach the available bandwidth. By alternating between
exponential and linear growth rates of \emph{cwnd} based on the backlog status,
SSthreshless Start eliminates the need for the Slow Start threshold,
\emph{ssthresh}, and the blind \emph{ssthresh} setting problem vanishes. The
use of backlog status at the bottleneck buffer also allows SSthreshless Start
to cope with various buffer sizes especially small buffer sizes that cause
performance degradation in many TCP startup variants.

Simulation results demonstrated that, compared with traditional Slow Start and
many other variants, SSthreshless Start significantly improves link utilization
during startup phase, meanwhile shows good performances to a wide range of
buffer size, propagation delay and bandwidth of bottleneck. Moreover, SSthreshless
Start shows good convergence behavior without adversely affecting coexisting
TCP connections. Therefore, being aware of backlog status, the enhanced throughput during startup phase is
achieved by using the bandwidth effectively and fairly rather than aggressively
depriving bandwidth from other co-existing TCP connections.

\end{document}